# Ovopub: Modular data publication with minimal provenance

Alison Callahan[1] and Michel Dumontier[1,2,3]

[1]Department of Biology, [2]Institute of Biochemistry, [3]School of Computer Science, Carleton University, Ottawa, Canada

**Abstract.** With the growth of the Semantic Web as a medium for creating, consuming, mashing up and republishing data, our ability to trace any statement(s) back to their origin is becoming ever more important. Several approaches have now been proposed to associate statements with provenance, with multiple applications in data publication, attribution and argumentation. Here, we describe the ovopub, a modular model for data publication that enables encapsulation, aggregation, integrity checking, and selective-source query answering. We describe the ovopub RDF specification, key design patterns and their application in the publication and referral to data in the life sciences.

## 1 Data publication and attribution: A simple problem with a (so far) complicated solution

The Semantic Web has unleashed a new paradigm for supporting a virtuous circle of data creators and data consumers where consumers can themselves act as creators with the ability to re-mix, re-phrase and re-publish content. However, these activities raise important questions about the provenance of any given data: What is it? Who made it? Where did it come from? How was it made? When was it made? What license governs its use, and can I reuse, modify or sell it? From the scientific research perspective, data publication may be incentivized by the potential for reuse provided attribution [1]. Indeed, our trust of data depends on being able to uncover and assess its provenance.

The Web of Data initiative is focused on developing practices to support the unique naming of individual resources on the Web [2] and to describe their provenance [3, 4][1]. However, how do we name, describe, publish and refer to assertions? Several efforts to address this challenge have so far been concerned with shrinking the scale of attributable and publishable objects from entire documents to some subset thereof. Nanopublications were developed to describe minimal assertions [5] between concepts, defined by [6] as the "smallest, unambiguous unit of thought". In its original formulation, a nanopublication consisted of two RDF named graphs: one containing a statement (an RDF triple) and another containing the annotations about that statement. Building on nanopublications, microattributions [7] aim to incentivize data contribution by recognizing the source of statements included in data resources, and has been used in describing gene variant data [8]. The nanopublications model has been recently updated [9] to use three named graphs: one for the statement(s), one for supporting statement(s) and one for related provenance. While certainly appealing on the surface, it is unclear what constitutes "support" in a nanopublication and whether supporting graphs can be exactly another's assertion graph in a different context. Similarly, while the micropublication model [10] uses a graph-based formalism of variable size and structure to construct an argumentation network linking textual statements and data as evidence for claims, it is unclear how a statement differs from a claim. These contributions are indeed an important step forward, but do not address the larger, more inclusive question: How can we describe self-contained units of knowledge, of *any* size or level of complexity, such that they can be published, shared, reused, extracted, modified, and republished? Moreover, neither model has yet been demonstrated for selective fact-based information retrieval across potentially billions of similar statements in which the structure and irregularity of content may yet pose significant challenges.

[1]http://www.w3.org/TR/prov-overview/

Here we propose the ovopub as a simpler, but effective knowledge publication model for describing any set or sequence of statements along with their provenance. We use ovo - from the Latin *ab ovo* - to refer to the earliest possible point in time at which an assertion could be described. The ovopub is simple by design and may be applied to represent knowledge of varying complexity and size. We posit that the ovopub is sufficient to describe any kind of statement and make it publishable and attributable, while its simple structure will facilitate widespread deployment to create, link, share and query ovopub networks. Specifically, the ovopub (i) embodies the simplest structure necessary to describe data, its provenance and digital rights, (ii) enables construction of more complex statements and arguments, (iii) allows encapsulation of selected statements, (iv) facilitates source restricted information retrieval and (v) ensures integrity of published data.

## 2 The Ovopub: Linking statements with provenance

An ovopub *contains* and *links* to one or more resources and/or statements, including those describing its provenance (**Fig. 1**). A resource is an object identified by a unique identifier. A statement is an n-tuple that either a) describes a relationship between two resources or b) assigns a literal value to a resource. There are three basic kinds of ovopubs: assertions, records and collections. An *assertion ovopub* contains a single statement that may be true or false. A *record ovopub* contains a single fully connected network of statements. A *collection* contains one or more resources and/or ovopubs, and is specifically meant to share or restrict a search to the resources contained therein. Essential ovopub provenance includes the creator(s) of the ovopub, a timestamp of when the ovopub was created, and a license to specify the rights and responsibilities of the creator and user of the ovopub. The integrity of any ovopub may be validated against a computed hash that is associated with, but external to the ovopub (not further described here).

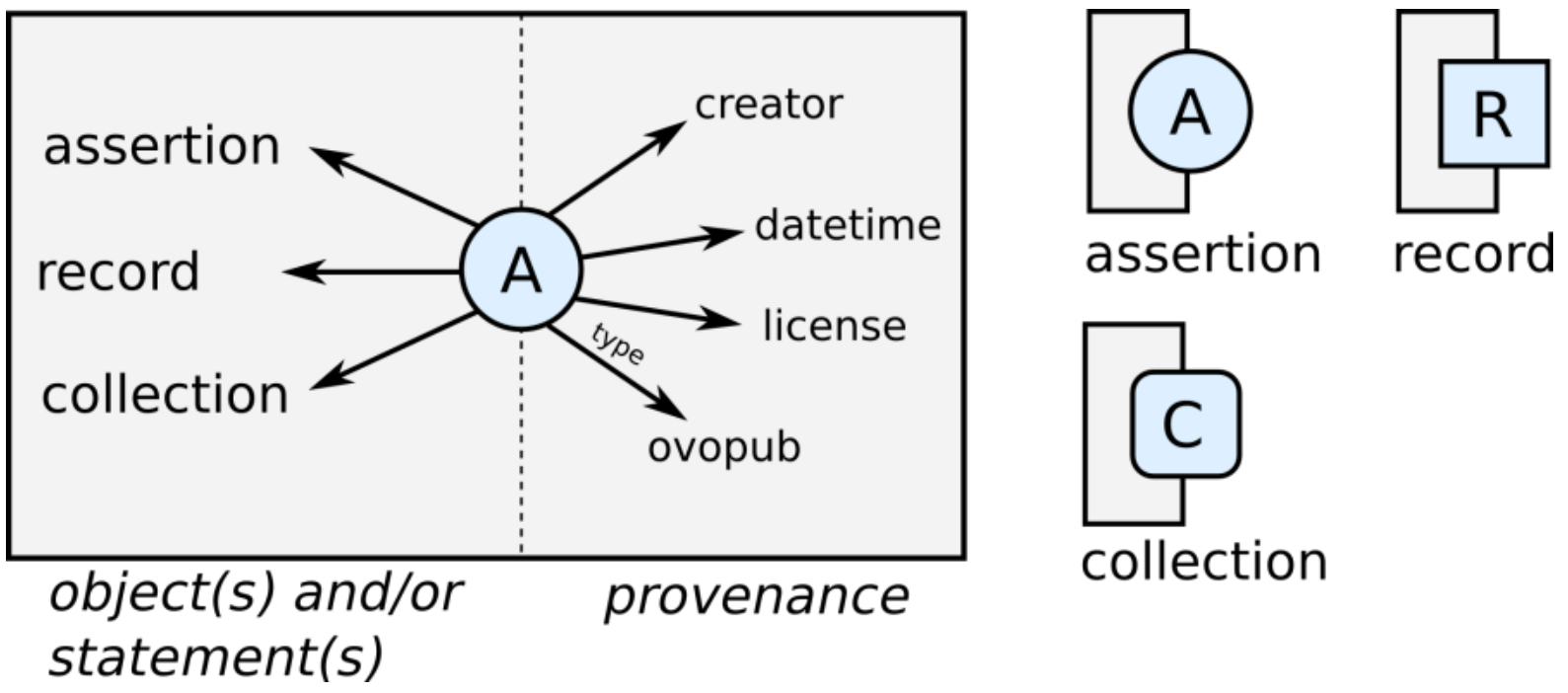

**Fig. 1.** The basic structure of an ovopub. An ovopub 'A' contains and links to its data (left) and its provenance (right). We use three symbols throughout to distinguish the assertion ovopub, record ovopub and collection ovopub.

## 2.1 Patterns for building and extending ovopub networks

### 2.1.1 **The Chaining Pattern.**

Ovopubs may be chained together through new assertions in a manner that both retains the original provenance of each ovopub and establishes the provenance of the established chain(s). **Fig. 2** illustrates such a case where three resources E, F and G are linked together in a record ovopub A. Assertion ovopub B describes the relationship P1 between E and F, and assertion ovopub C describes the relationship P2 between F and G. Each ovopub may have been created by different sources and/or at different times, and this information can be captured in their respective ovopub-linked provenance. Assertion ovopub D establishes that A is related-to B (we use *related-to* as a generic example predicate; any more appropriate predicate may be used), while collection ovopub E contains ovopubs A, B, C and D together for reference and potential future reuse. Importantly, a collection ovopub like ovopub E may be invoked in order to isolate and specifically refer to a set of selected resources or ovopubs extracted from a potentially large and complex network.

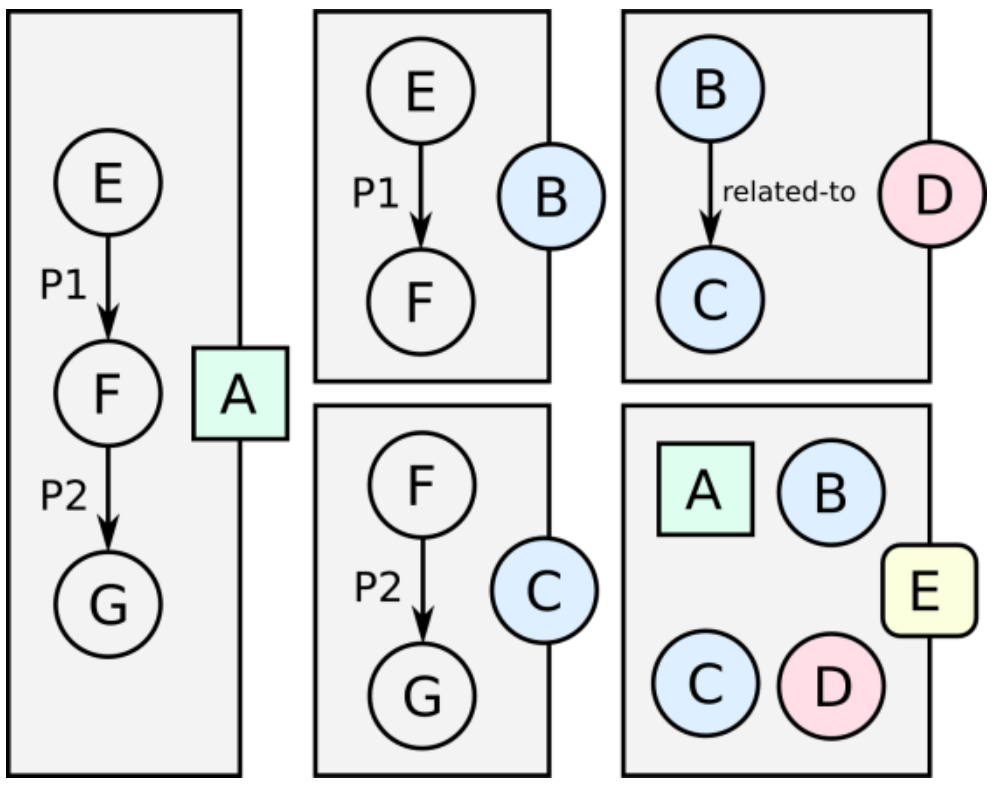

**Fig. 2**. Chaining statements together using a combination of assertion, record and collection ovopubs. A record ovopub A contains the statements linking resources E, F and G. These statements can be described in individual assertion ovopubs B and C, and their relationship described in a another assertion ovopub D. These record and assertion ovopubs can be grouped into collection ovopub E.

The chaining pattern has innumerable potential uses, including but not limited to (i) providing a rich historical provenance of an object or assertion [11-13] (*e.g.* statement A has-source statement B), (ii) establishing argumentation [10] (*e.g.* statement A supported-by/disputed-by statement B) and (iii) making ontological assertions [14] (*e.g.* axiom A equivalent-to axiom B). These specific cases differ from the general use case only by the predicate linking ovopubs A and B - the ovopub graph structures remain the same. In every case, the provenance for any ovopub is captured as part of the ovopub itself. However, should the ovopub creator decide to include an explicit provenance of the statement itself, another ovopub would link the first ovopub URI to its source via some appropriate provenance relation.

**The Aggregation Pattern.**

Given the potential for redundancy of assertions in an ovopub network, it becomes necessary to aggregate statements based on identity in the non-provenance content of an ovopub, or some other criteria of identity or similarity. **Fig. 3** shows how ovopubs can be used to aggregate statements. The ovopub aggregation pattern is broader than that of the nanopublication cardinal assertion, where aggregation occurs over those assertions that contain exactly the same subject, predicate and object [6, 15]. Aggregation ovopubs may be gathered into a collection ovopub for reference and reuse.

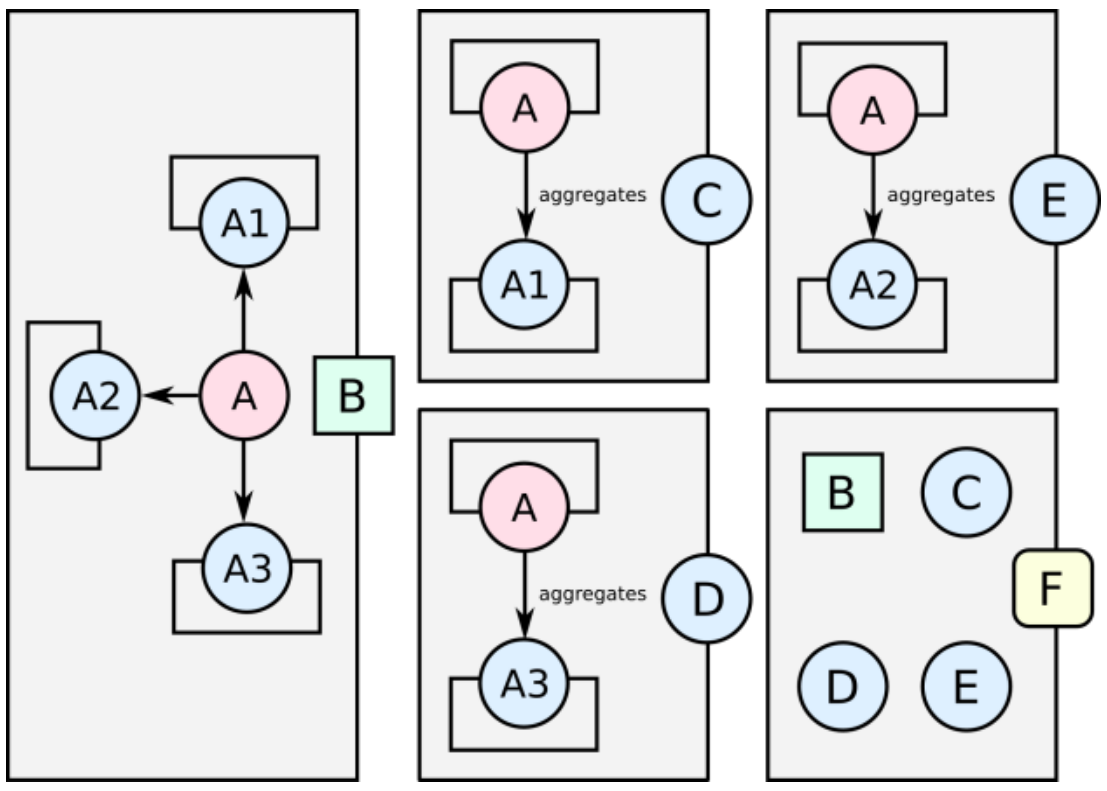

**Fig. 3.** Aggregating statements into a collection ovopub. A single record ovopub B contains an assertion ovopub and the three ovopubs it aggregates. These can be individually described in assertion ovopubs C, D and E. All record and assertion ovopubs can be grouped into collection ovopub F.

### 2.2 Ovopub RDF specification

Ovopubs that are represented using the Resource Description Framework (RDF) must implement the following specification in order to be considered a valid ovopub. An ovopub is a named graph that contains and explicitly links to statements of interest. Thus, all statements must be expressed as quads such that the referent graph is the ovopub URI. We use four vocabularies to describe types and relations: RDF (rdf), RDF schema (rdfs), XML Schema (xsd), Dublic Core Metadata Terms (dc), and the Semanticscience Integrated Ontology (sio). Assertion ovopubs (**Fig. 4A**) must be typed as sio:assertion-ovopub[2] and be explicitly reified to the referent subject, predicate and object of the triple using the rdf:subject, rdf:predicate and rdf:object relations, respectively. A record ovopub (**Fig. 4B**) must be typed as a sio:record-ovopub[3] and linked to each statement with the rdfs:member relation. A collection ovopub (**Fig. 4C**) must be typed as a sio:collection-ovopub[4] and be explicitly linked to its member ovopubs using rdfs:member. The sio:assertion-ovopub, sio:record-ovopub and sio:collection-ovopub are all subclasses of sio:ovopub[5]. Ovopubs may optionally use rdfs:label to specify a title (with appropriate language tag), dc:identifier to specify a source specified non-URI identifier, dc:description to provide a more detailed description (with language tag). The ovopub creator must be specified using dc:creator linked to either a literal specified as an xsd:string or to a resolvable linked data URI (*e.g.* a FOAF entry). The ovopub date of creation must be specified as a timestamp using dc:date as the datatype property and the literal value a xsd:datetime. The ovopub license must be specified using dc:rights with the value specified as a URI pointing to the license document. We recommend using the Creative Commons by attribution (CC-BY) licenses for data.

---

2 http://semanticscience.org/resource/SIO_001301
3 http://semanticscience.org/resource/SIO_001302
4 http://semanticscience.org/resource/SIO_001303
5 http://semanticscience.org/resource/SIO_001300

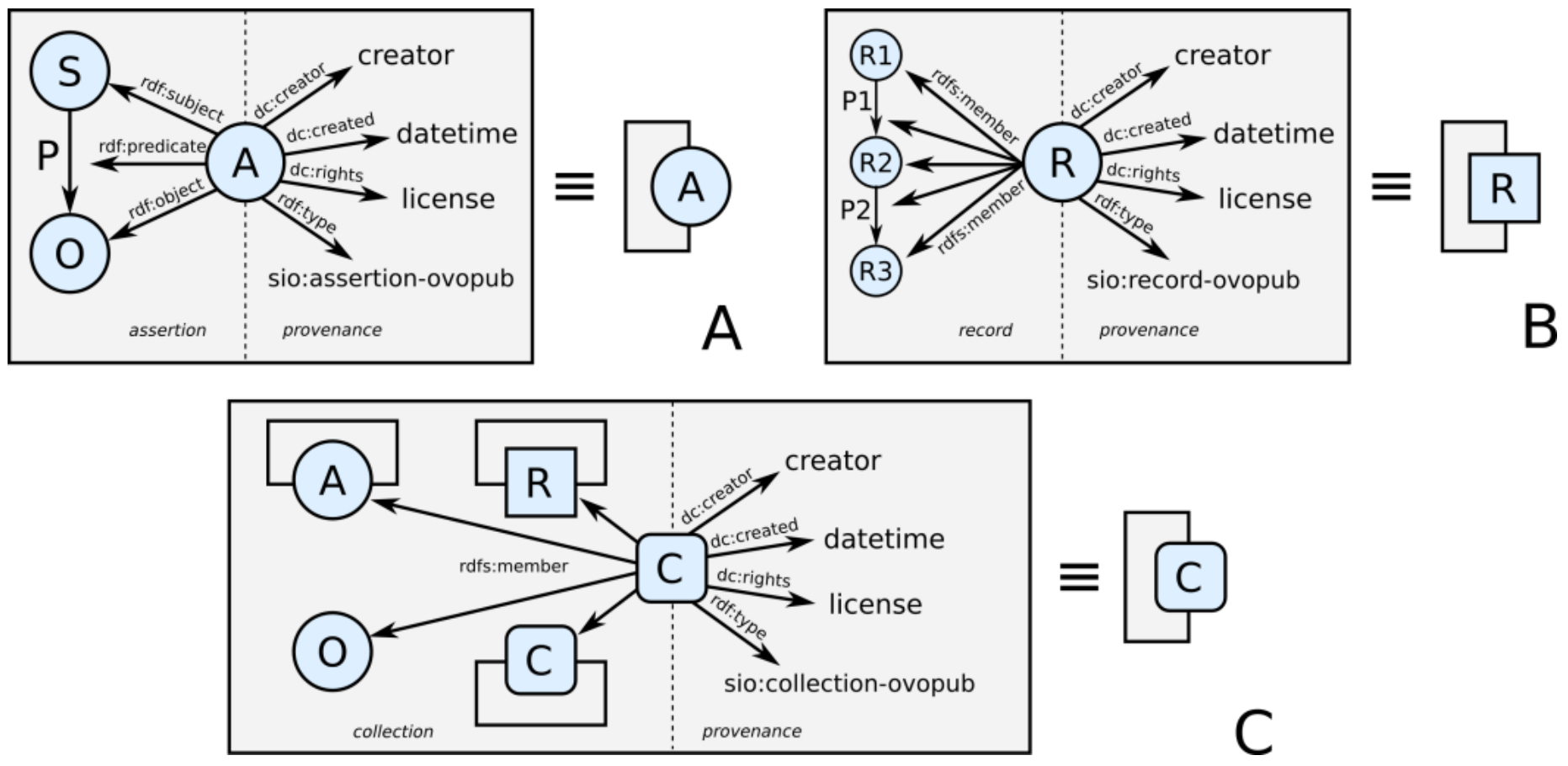

**Fig. 4.** Assertion (A), record (B) and collection (C) ovopub RDF specifications. An assertion ovopub contains a reified statement and links to each element using rdf:subject, rdf:predicate and rdf:object. A record ovopub contains and links to a set of statements using the rdfs:member predicate. A collection ovopub contains and links to any resource (*e.g.* object, predicate, assertion ovopub, record ovopub or collection ovopub) using the rdfs:member predicate. Each ovopub type is specified using the rdf:type relation. Provenance details are l using relations from the Dublin Core (DC) vocabulary.

## 3 Modular knowledge representation with ovopubs

We illustrate the application of the ovopub model to a complex scenario: an entry in the iRefIndex database of protein-protein interactions (PPIs). iRefIndex [16] collects interactions reported in 13 source databases and groups interactions based on taxon and sequence identity as well as sequence similarity of its molecular participants. In this fashion, iRefIndex serves as a natural aggregation point and must therefore distinguish the relations that were asserted by the source database and the relations that it asserts for the purpose of data aggregation. **Fig. 5** illustrates the capture of relations (i) between an individual PPI (BioGRID:464511), the versioned iRefIndex record, and the dataset of which it is a part, (ii) the description of interactions in terms of its protein participants, method of PPI identification and publication, and (iii) the aggregation of PPIs into interaction groups based on the aggregation criteria set out by iRefIndex. This demonstrates the utility of ovopubs for (i) encapsulation and versioning, (ii) description of domain knowledge and (iii) aggregation and information reduction. Ovopubs, and thus their provenance, can be as fine-grained as the simplest statement in a dataset. For example, the simplest ovopub in this case is the assertion describing the protein interaction tuple (**Fig. 5B**). One can describe the provenance for this extracted tuple, include it in any arbitrary dataset, and also link to the ovopub for the BioGRID record in order to explicitly describe its source (**Fig. 5B**).

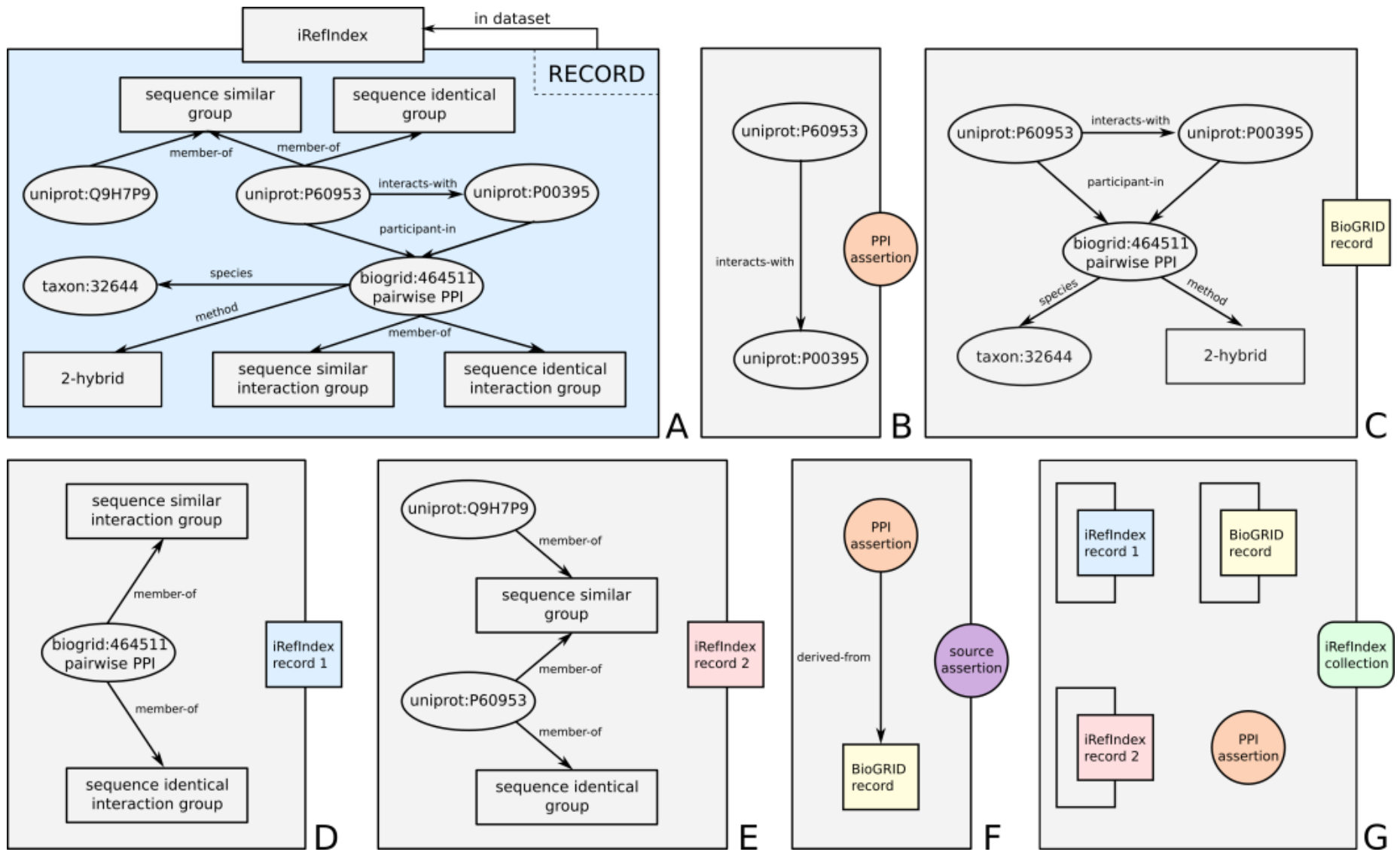

**Fig. 5.** Relations between data items in an iRefIndex record for BioGRID:464511 (A) and their corresponding representation as ovopubs. An assertion ovopub (B) describes the protein-protein interaction (PPI) relationship between two UniProt proteins. A record ovopub (C) describes this PPI as collected by BioGRID, while two additional record ovopubs (D, E) describe the membership of the BioGRID PPI and participating UniProt proteins in iRefIndex interaction and similarity groups, respectively. An assertion ovopub (F) links the PPI ovopub URI to the BioGRID record ovopub URI, to indicate its source. A collection ovopub (G) collects a subset of the preceding ovopubs.

# 4 Using ovopubs for context-sensitive information retrieval

Redundancy, ambiguity and contradiction are major issues in the naive integration of massive amounts of linked data. The ability to capture explicit provenance of an assertion through argumentation networks ensures that users can filter data according to original data providers, for example to include statements generated only by a trusted source. The ovopub model also enables users to create and refer to custom collections that include selected ovopubs, thereby potentially increasing the quality of public data. Yet an important issue remains - can we restrict a query to only the contents of selected data that are potentially entangled in a sea of assertions and arbitrarily constructed collections? Doing so would require the ability to compute a transitive closure, and to restrict the answer of a query to those resources and assertions that are only in the selected datasets. Using the Semantic Web framework, it is indeed possible to restrict a SPARQL query involving a transitive closure over a selected property to the set of ovopub collections and assertions of interest. SPARQL 1.1 offers property paths to construct transitive closures over the explicit reified relations (and their inverses) in the ovopub model (rdfs:member, rdf:subject, rdf:predicate, rdf:object). For this purpose, the object properties rdfs:member, rdf:subject, rdf:predicate and rdf:object can be asserted to be sub-properties of sio:is-transitively-related-to[6] and the set of ovopubs to be queried restricted using the SPARQL FROM keyword. Similarly, ovopub collections can be considered OWL ontologies in which the assertions can reasoned about using property chaining. Thus, while explicit reification of ovopubs is not necessary for querying assertions directly, it serves the important purpose of enabling complex, context-sensitive information retrieval.

## 5 Discussion

The ovopub is simple by design in order to maximize its reuse in innumerable scenarios. The ovopub is but a single object that contains and links to basic elements of data-centric provenance (what, who, when, rights) and its content, whether a simple assertion involving an object, another ovopub assertion, record or collection, or a collection of ovopub assertions, records and/or collections. The ovopub model enables encapsulation, linking and aggregation and facilitates complex queries that consider elements of provenance and trust.

The ovopub also establishes a strong base upon which to structure and enhance linked open data and to create networks of information pertaining to historical provenance and versioning, argumentation, text mining, ontology building, and other application areas. Massive open data integration efforts such as Bio2RDF [17, 18] could establish the ovopub as a finer grained model for assertion publishing and provenance and better describe dataset processing and provenance, perhaps drawing from recently developed vocabularies such as PAV [19] and PROV [4]. Furthermore, the ovopub creates new opportunities to develop ontologies to further describe different kinds of assertions (*e.g.* Gene Ontology).

The ovopub can be readily contrasted with nanopublication and micropublications. The ovopub is simpler as it consists of only a single named graph with key provenance information directly contained in and associated with the ovopub graph. In contrast, the nanopublication graph is explicitly linked by three relations to three named graphs: one that contains the statement(s), one that contains the supporting statement(s) and one that contains the related provenance. The ovopub therefore reduces the number of required statements by consolidating the statement and provenance graph into a single ovopub graph and leaving the supporting graph to be specified as another ovopub. It also removes the ambiguity of what should be included in the supporting statement graph, whether an assertion ovopub can be used as a supporting graph in another publication, and sidesteps the problem of how to manage a change in the supporting statement graph *vis-a-vis* the original nanopublication. In addition, unlike the nanopublication or the micropublication, the provenance of an ovopub is distinct from the provenance of the ovopub payload, whose historical provenance (who stated it? how was it generated? where was the statement obtained? *etc.*)

[6] http://semanticscience.org/resource/SIO_001247

could either be published a set of provenance-oriented assertion ovopubs or could directly stated in a record ovopub.

Cardinal assertions aggregate syntactically identical statements described in nanopublications to establish a confidence or evidence score [15]. Through the iRefIndex exemplar, we show that ovopubs can be used to aggregate syntactically identical, semantically equivalent, or semantically similar statements from multiple sources (each being described in an ovopub) from which evidential strengths may be computed if desired. Cardinal assertions can be easily computed through hash sums on the payload assertion or the specified members of the dataset as identified through relations provided by explicit reification. More importantly, the integrity of an ovopub can also be assessed. In this case, the entirety of the ovopub would be subject to a hash sum whose value could be recorded in a separate ovopub signed by the creator. In this way, data 'hijacking' by adding additional links or statements to an ovopub after it has been created and published could be detected.

With the availability of increasingly powerful triple stores and reasoners, adopting the ovopub model for existing large linked data networks is certainly feasible. OpenLink Virtuoso and Ontotext's OWLIM provide cluster-capable triple store implementations, while the Digital Enterprise Research Institute at the National University of Ireland Galway with Fujitsu Laboratories have announced a new linked data platform capable of querying billions of triples at greatly improved speeds [20]. Similarly, the Web-scale Parallel Inference Engine (WebPIE) enables large scale reasoning [21].

## 6 Summary

The ovopub is a new model for data publication. Its simplicity and modular design support the creation of networks of arbitrary size and complexity. We have described the minimal elements required for an ovopub, as well as how ovopubs can be used to make assertions about objects, literals, statements or even collections of statements. We have described the application of ovopubs to address the requirements of description, encapsulation, aggregation, data integrity and selective-source query answering. Our next steps are to implement the ovopub model for Bio2RDF datasets, evaluate aspects related to performance and scalability, and to explore the use of ovopub networks for knowledge discovery.

## 7 Acknowledgements

This research was supported in part by an Ontario Ministry of Research and Innovation Early Researcher Award to MD and an NSERC Canada Graduate Scholarship – Doctoral to AC.